# Coarse-Graining with Equivariant Neural Networks: A Path Towards Accurate and Data-Efficient Models

Timothy D. Loose,[#] Patrick G. Sahrmann,[#] Thomas S. Qu, and Gregory A. Voth*

Department of Chemistry, Chicago Center for Theoretical Chemistry, James Franck Institute, and Institute for Biophysical Dynamics, The University of Chicago, Chicago, IL 60637, USA

[#]Authors contributed equally

*Corresponding Author: gavoth@uchicago.edu

## Abstract

Machine learning has recently entered into the mainstream of coarse-grained (CG) molecular modeling and simulation. While a variety of methods for incorporating deep learning into these models exist, many of them involve training neural networks to act directly as the CG force field. This has several benefits, the most significant of which is accuracy. Neural networks can inherently incorporate multi-body effects during the calculation of CG forces, and a well-trained neural network force field outperforms pairwise basis sets generated from essentially any methodology. However, this comes at a significant cost. First, these models are typically slower than pairwise force fields even when accounting for specialized hardware which accelerates the training and integration of such networks. The second, and the focus of this paper, is the need for the considerable amount of data needed to train such force fields. It is common to use 10s of microseconds of molecular dynamics data to train a single CG model, which approaches the point of eliminating the CG model's usefulness in the first place. As we investigate in this work, this "data-hunger" trap from neural networks for predicting molecular energies and forces can be remediated in part by incorporating equivariant convolutional operations. We demonstrate that for CG water, networks which incorporate equivariant convolutional operations can produce functional models using datasets as small as a single frame of reference data, while networks without these operations cannot.

# 1. Introduction

Molecular dynamics (MD) has proven to be a powerful tool to study the molecular underpinnings behind a variety of interesting phenomena in the biological and material sciences.[1-3] The efficient integration of Newton's equation of motion in MD provides a fast method by which statistical information of a molecular system can be averaged over to arrive at conclusions about the thermodynamics and dynamics of the system. A variety of MD techniques have been developed to investigate systems at a range of different accuracies at the all-atom (AA) level. At the other end of the spectrum is coarse-graining (CG), which seeks to accurately simulate a molecular system at a resolution below that of atomistic MD.[4-7] In recent years there has been a large amount of interest in applications of machine learning (ML) to molecular simulations both at the AA (and *ab initio*)[8-16] and CG[17-21] levels. While some of these approaches apply ML to generate a pairwise CG force field (see, e.g., the original work by Voth and co-workers[22-25]), most work has pursued the idea to treat the ML model – typically a deep neural network (DNN) – as the CG force field itself. On the *ab initio* MD end, DNN force fields tend to integrate much faster than full quantum treatments of the atoms as they can typically be evaluated as a series of matrix multiplications and thus may be useful as a method to speed up integration while maintaining an acceptable level of accuracy. On the CG end, however, ML-based methods tend to be slower than a simple pairwise CG forcefield, although they can be more accurate as they can naturally incorporate many-body correlations. The DNN can also be better at fitting the interaction data than the linear regression[23, 24, 26] or relative entropy minimization[27, 28] (REM) CG'ing-based methods that are more typically employed in "bottom-up" CG model development.

ML has a long history of drawing inspiration from nature to create powerful models which can tackle problems that were previously considered intractable. The architecture of the first neural networks was, as their name suggests, inspired by neural function in the brain.[29] Similarly, convolutional neural networks take advantage of processes found in animal eyes to identify and classify features above the single pixel-level in image processing.[30, 31] Both the brain and eye are remarkably powerful tools for learning and image processing respectively, so it is perhaps no surprise that these models can succeed at such tasks. In the application of DNNs to molecular systems, it should also come as no surprise that taking inspiration from physics can lead to better results. Most successful DNN-based methods incorporate physical constraints into their architectures and training schemes to improve the often-nebulous connection between DNN regression and physical reality as well as to speed up training. Typical DNN based force fields, such as CGnet[32] and CGSchnet[33], utilize the same objective function as the multiscale coarse-graining (MS-CG) method[22-24], the mean-squared error between the forces of the mapped reference data and the CG model.[23, 24] The difference lies in the usage of a neural network to learn these forces versus the force matching approach which employs least squares regression over a simpler set of model parameters, typically either Lennard Jones parameters or B-splines (or both).

Further physical intuition can be applied to the structure of the networks themselves. The DNNs work in two stages. The first is a featurization stage in which the raw Cartesian coordinates for each particle in a configuration are converted into more natural internal coordinates while the second is a neural network that learns particle-wise energies and forces from these featurized configurations. The simplest featurization scheme corresponds to converting the coordinates into interparticle distances or their inverses as well as particle types, which can be done directly as is the case for CGnets.[17] These features are then subjected to physically inspired energy priors, i.e., harmonic for bonded particles and repulsive for non-bonded ones. This frees the energy-predicting network from needing to learn those features of the CG Hamiltonian and allows it to learn corrections to the priors instead.

Another approach for featurizing molecular configurations which can generate more accurate results is to embed these features into a graph neural network as in Schnets and CGSchnets[8, 18] which are naturally suited to representing molecular systems. Each node of the graph represents a CG site, and each edge a distance between the two CG sites representing each node. Convolutions over these graph elements can be performed analogously to convolutions over pixels in 2-D images, giving graph neural networks a powerful tool to pool information across a variety of spatial scales.[34] These networks can then be trained to learn an effective embedding of the CG configuration which optimally predicts CG forces and energies, improving over the set of hand-selected internal coordinates used by CGnets. This embedding network fits into the previously discussed architecture in between the original featurization into internal coordinates and before the energy prediction network. This method also produces networks that may be more inherently transferable, as the embedding network can develop an effective embedding for any configuration so long as there are no CG types that have not been seen by the network. For systems such as proteins, this is possible to accomplish, so long as the training dataset contains all 20 amino acids.

Both the handpicked featurization and the graph neural network implementation ensure that all resulting CG features are at least invariant to rotation and translation, and the overall force-field is equivariant. Equivariant neural networks specifically guarantee the proper transformation behavior of a physical system under coordinate changes. This relationship can be described more precisely in group theoretic terms, in which a group $G$ operates on vector spaces $X$ and $Y$. A function $f(x)$ that maps from $X$ to $Y$ is equivariant with respect to $G$ if

$$D_Y[g]f(x) = f(D_X[g]x) \qquad (1)$$

where $D_X[g]$ and $D_Y[g]$ are representations of element $g$ in the vector spaces of $X$ and $Y$ respectively.

Recently, a class of equivariant neural networks has been developed for atomistic molecular systems which incorporates equivariance into the hidden layers of the network.[10, 11, 15] These methods incorporate full vector information of the relative positions of atoms in addition to higher order tensor information to guarantee that the magnitude of the forces produced by these networks are invariant to rotation, translation and reflection (also known as the Euclidian or E(3) symmetry group) while the unit vectors describing the directions of these forces are equivariant under these operations. These properties impose a restraint on the networks based on the physics which in theory should make the networks far more capable of representing and predicting molecular forces. Specific implementations of equivariance differ from architecture to architecture, but the present work focuses on the NequIP[11] and Allegro models. These architectures take advantage of the natural translational and permutational equivariance of the convolution, and enforce that the convolutional filters are products of radial functions and spherical harmonics, which are rotationally invariant to achieve full E(3) equivariance:[35]

$$S_m^{(l)}(\overrightarrow{r_{ij}}) = R(r_{ij})Y_m^{(l)}(\hat{r}_{ij}) \quad (2)$$

where $S_m^{(l)}(\overrightarrow{r_{ij}})$ is a convolutional filter over full distance vectors between atoms, $r_{ij}$ is the scalar distance associated with $\overrightarrow{r_{ij}}$, and $\hat{r}_{ij}$ is the corresponding unit vector. Allegro and NequIP differ in that NequIP is globally equivariant. NequIP achieves global equivariance via a message-passing layer which passes messages from adjacent graph nodes. These layers can learn a variety of functions, from graph convolutions to graph-wide targets, which encode information about the entire system.[36] Allegro removes this message passing layer and only achieves local equivariance as a cost to allow for parallelization of network evaluation, thereby allowing it to scale to much larger systems.

The results of these equivariant networks can address a key weakness of ML: the large dataset requirements for training neural networks. To generate an effective DNN, one must supply a very large amount of training data (usually MD frames in the case of bottom-up CG model development), which can in turn make the resulting CG model less useful (or potentially useless) as it inflates the time required to calculate results. For example, prior training of a CGSchnet architecture to produce a force field for chignolin, a mini-protein containing 10 amino acids, required 180 microseconds of reference simulation.[18] By contrast, a more conventional CG hetero-elastic network model[37] for full-length integrin, containing 1780 residues, was generated using 0.1 microseconds of MD simulation.[38] This disparity may call into question the usefulness of DNNs as CG force fields under certain circumstances when applied to lower resolution CG models even considering the highly accurate results they may generate. As it turns out and will be shown in this paper below, one primary reason for the large number of examples (data) required to train DNN force fields is the equivariance with respect to molecular forces. Rotation invariant DNN-based methods must learn the equivariance of forces via training reinforcement, which adds considerably to the data cost of creating these models. Equivariant neural networks, on the other hand, build this information into the model inherently and have been shown to predict interatomic energies and forces for small molecules at atomic resolution with three orders of magnitude less training data than symmetry invariant architectures and to do so with even greater accuracy.[11]

While training neural networks to predict energies and forces of atomistic resolution systems from *ab-initio* quantum data is not the same as training a CG model from atomistic data, there is a natural analogy of learning to predict forces at a lower resolution from higher resolution data. In the former case, the high resolution is the quantum description of the system, while the low resolution is the atomistic description. In the case of CG'ing, the high resolution model is the atomistic description, while the low resolution is some chosen CG lower resolution. A key difference is that most methods that learn atomistic descriptions from quantum data treat bonded and non-bonded interactions as the same, and rely purely on internal coordinates to decide this, while CG DNN methods use labels and alternative energy priors to do this. This is necessary for complex CG systems as bond breaking and forming are typically ignored for these models, and because the length scales of bonded interactions can easily match and overlap with that of the non-bonded ones. For this reason, the currently available equivariant neural network-based methods must be used carefully when applying them to CG systems.

There are certain cases in which the methods are fundamentally identical. The simplest case is that in which there are no bonds whatsoever, and each CG site or "bead" corresponds to an entire molecule. In this case, the act of making a CG model is equivalent to the act of reducing the quantum description of a nonreactive single particle, such as helium, to its atomistic representation. For this reason, the work in this paper is limited in scope to the coarse-graining of single-site liquids, namely single-site water as a key example. Water is also an ideal test case for a DNN CG method due to the high levels of correlation caused by the underlying hydrogen bonding. For this reason, single-site CG water models tend to fail to predict proper center of mass radial distribution functions (RDFs) for water unless they incorporate many-body correlations.[39-43]

In this work we present an analysis of DNN-based CG models of single-site CG water utilizing invariant and equivariant convolutional operations. For the invariant model, the Deep Potential Molecular Dynamics method with smoothed embedding (DeePMD)[44, 45] is utilized. For the equivariant model, the Allegro model[46] is utilized. Each method is applied to water in the limit of low sampling: a maximum of 100 consecutive MD frames are used to train each model.

This remainder of this paper is organized out as follows: First, a discussion of the methods is given, with the hyperparameters for all ML methods as well as all MD simulation parameters. A discussion of DeePMD and Allegro models is also presented. Following this, results for each model are presented. Pairwise RDFs are analyzed and compared to mapped atomistic reference data. Three-body angular correlations are also analyzed. Finally, the stability of each force field in the low sampling limit is discussed as well. These results are then discussed and conclusions on the usefulness of equivariant particle embedding in the field of CG modeling are drawn.

## 2. Methods

In order to generate the dataset used to train the models, the LAMMPS[47] and GROMACS[48] MD programs were used to simulate 512 TIP3P[49] and SPC/E[50] water molecules for a total of 10 nanoseconds in the constant NVT ensemble, respectively. For both models, a Nosé-Hoover thermostat[51, 52] was used to maintain the simulation at 300 K, and frames were captured every 2 ps for a total of 5000 frames, though far fewer were used in the training of the Allegro and DeePMD models. The resulting trajectory was mapped to a 1 CG site per water resolution using a center of mass (COM) mapping scheme. This was then passed as a training dataset to DeePMD and Allegro.

The DeePMD method consists of both an embedding network and a fitting network. The embedding takes pairwise distances as input and outputs a set of symmetry invariant features which include three-body information such as angular and radial features from nearby atoms, denoted by the authors as the se_e2_a embedding. Notably, this embedding network is not a graph neural network. However, before the interatomic distances are fed into this matrix, they are converted into a set of coordinates based on inverse distances:

$$\{x_{ij}, y_{ij}, z_{ij}\} \to \{s(r_{ij}), \hat{x}_{ij}, \hat{y}_{ij}, \hat{z}_{ij},\} \qquad (3)$$

$$s(r_{ij}) = \begin{cases} \dfrac{1}{r_{ij}}, & r_{ij} < r_{c1} \\ \dfrac{1}{r_{ij}}\left\{\dfrac{1}{2}\cos\left[\dfrac{\pi(r_{ij}-r_{c1})}{(r_{c2}-r_{c1})}\right]+\dfrac{1}{2}\right\}, & r_{c1} < r_{ij} < r_{c2} \\ 0, & r_{ij} > r_c \end{cases} \qquad (4)$$

where $x_{ij}$, $y_{ij}$ and $z_{ij}$ refer to the x, y and z projections of $r_{ij}$ the distance between two particles i and j, and $\hat{x}_{ij} = \dfrac{s(r_{ij})x_{ij}}{r_{ij}}$, $\hat{y}_{ij} = \dfrac{s(r_{ij})y_{ij}}{r_{ij}}$, and $\hat{z}_{ij} = \dfrac{s(r_{ij})z_{ij}}{r_{ij}}$. This set of features is then converted via the embedding network into a matrix of features that preserves the rotational, translational, and permutational symmetry of the system. These features are passed through per-atom subnetworks which compute the energy contribution from each atom to the total system energy. The gradients of these per-atom energies can then be used to calculate interatomic forces during an MD simulation. During training, the DeePMD model sees individual atoms as training samples which can be batched as usual.[45]

Allegro is an extension of the NequIP model which trades global equivariance gained via a message-passing graph neural network for local equivariance in order to provide much greater scaling capability.[46] In Allegro, two sets of features are generated by the initial featurization for each pair of particles. The first is a scalar set of features which consist of interatomic distances and labels for each chemical species in the interaction. This feature set is symmetry invariant, as in the case of DeePMD. The second feature set contains unit vector information that correspond to these interatomic distances which are projected onto spherical harmonic functions. These features are then embedded through a series of layers onto a new equivariant feature set which is then fed into a multilayer perceptron (MLP) which predicts the energy of the interaction. The total energy of the system can be calculated as the sum of these energies, and the forces can be calculated via the gradients of these energies.

For both Allegro and DeePMD models, a common network size was selected to ensure that differences in the performance of the models were most strongly correlated with the number of training samples. The embedding networks were composed of three layers with widths [8, 16, 32]. In the case of DeePMD models, this format is converted into a ResNet,[53] for which no timestep was selected. The energy fitting networks were also composed of three layers each with widths [32, 32, 32]. Each network was given a maximum cutoff for the environment of each atom of 7 Angstroms. Training parameters such as the number of epochs, learning rates, and early stopping were left up to the defaults of each model archetype to ensure that each model was trained according to its normal usage. Specific parameters for each model may be found in the Supplementary Information in the form of DeePMD and Allegro input files.

A total of 4 models were trained according to the preceding description. For DeePMD, two models were trained, one using 100 frames (or 200 ps) of training data, and one using 10 frames (or 20 ps) of data. Two Allegro models were trained using 100 frames and 1 frame of training data. The TIP3P CG water models were simulated for 2,500,000 timesteps and the SPC/E CG water models were simulated for 1,000,000 timesteps. We note that the models presented here are trained in the data scarcity limit, and that modern equivariant neural networks are practically trained on orders of magnitude larger datasets than the datasets considered here. Each model was tested on a simulation of 3916 water molecules using a Nose-Hoover thermostat in the constant NVT ensemble at 300 K just as was the reference data. All models, except for the 10-frame DeePMD model, were simulated using a timestep of 2 fs, while the 10-frame DeePMD model used a 0.5 fs timestep, a choice which is explained in the Results and Discussion section.

To calculate the RDFs, an outer cutoff for each model was selected to be 10 Angstroms. The 3-body angular distributions, $P(\theta)$, were calculated for water using the following equation:

$$P(\theta) = \frac{1}{N} \left\langle \sum_I \sum_{J \neq I} \sum_{K > J} \delta(\theta - \theta_{JIK}) \right\rangle_{R < R_C} \tag{5}$$

where $N$ is a normalization constant equal to the largest value in the calculated sum and $R_c$ is the cutoff radius. For these correlation functions, an outer cutoff of 4.5 Angstroms was selected which corresponds with the second solvation shell of water originating from its tetrahedral ordering.[54] The 3-body correlations were calculated between 30 and 150 degrees with a bin width of 1 degree, which captures the full extent of the 3-body correlations seen in tetrahedral water.[42]

## 3. Results and Discussion

*Force Error.* The validation RMSE for each ML model trained on the TIP3P and SPC/E water models are reported in Tables 1 and 2, respectively. For both water models, the 100-frame Allegro model demonstrates the best performance. Intriguingly, the performance of Allegro on a singular frame is identical to the performance of DeePMD on two orders of magnitude more data for the TIP3P water model, and is similar in magnitude for the SPC/E water model. The DeePMD 10-frame exhibits the largest validation RMSE, and as demonstrated from simulation, deviates the furthest in capturing the structural correlations in the reference models.

| Model | RMSE (kcal/mol Å) |
|---|---|

| Model | RMSE |
|---|---|
| 10-frame DeePMD | 5.81 |
| 100-frame DeePMD | 3.66 |
| 1-frame Allegro | 3.66 |
| 100-frame Allegro | 3.55 |

**Table 1.** RMSE values of ML models trained on the TIP3P water model.

| Model | RMSE (kcal/mol Å) |
|---|---|
| 10-frame DeePMD | 11.8 |
| 100-frame DeePMD | 3.54 |
| 1-frame Allegro | 3.64 |
| 100-frame Allegro | 3.49 |

**Table 2.** RMSE values of ML models trained on the SPC/E water model.

*Simulation stability.* Every model trained except for the 10-frame DeePMD model of TIP3P water was stable when simulated using a 2 fs timestep. The 10-frame DeePMD model suffered from severe energy drift at this time-step and the system quickly falls out of a liquid state.

a)

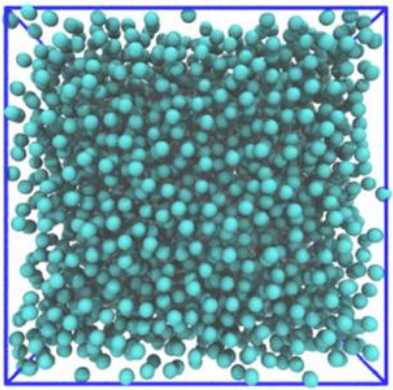

b)

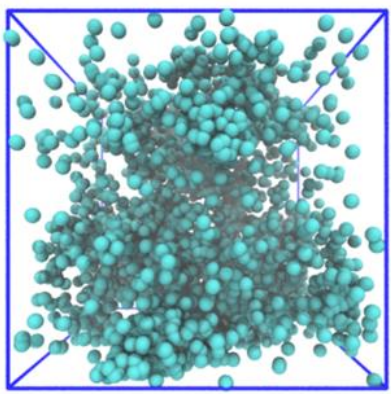

Figure 1. Post-equilibration snapshot of (a) the Allegro CG water model trained on 1 frame of AA reference data and b) the DeePMD water model trained on 10 frames of AA reference data of TIP3P water.

Figure 1 shows sample coordinates for the 10-frame DeePMD and 1-frame Allegro TIP3P models, detailing the extent of the instability of the former system. Interestingly, even the repulsive priors present in the DeePMD architecture cannot prevent the particles reaching unphysically close distances. Within 2,500 timesteps, the water sites coalesce into small clumps. To simulate the model long enough to collect data, a 0.5 fs time-step was selected, as a 2 fs timestep lead to an almost immediate loss of multiple CG particles from the simulation box. Furthermore, a 500 fs Nose-Hoover damping coefficient was not strong enough to stabilize the system at 300 K and the system exhibited a lower temperature throughout the simulation. On the other hand, all Allegro models performed stably and did not naturally tend towards a lower temperature or a collapsed state.

*Radial distribution functions (RDFs).* Figures 2 and 3 show RDFs for each CG model compared to that of the mapped reference system for TIP3P and SPC/E water, respectively. Each stable NN model does a good job of capturing the structural correlations of liquid water, a difficult task for a single site CG water model based on pair-wise non-bonded CG potentials. While none of the CG models developed from either AA model fully capture the depth of the well directly beyond the first peak, all models aside from the 1-frame Allegro model are close, with the 100-frame DeePMD model performing the best for both water models. The 1-frame Allegro models overestimate the height of the first peak, while the well depth is overestimated for TIP3P water and underestimated for SPC/E water. None of the models can capture the small peak around 6 Angstroms for TIP3P water and 4.5 Angstroms for SPC/E water, with the 100 frame models performing slightly better than the 1-frame Allegro models. The most noteworthy differences come from the 10-frame DeePMD model, which deviates significantly because of its propensity to collapse individual water molecules onto one another in the TIP3P model, and its poor representation of the underlying liquid structure for the SPC/E model.

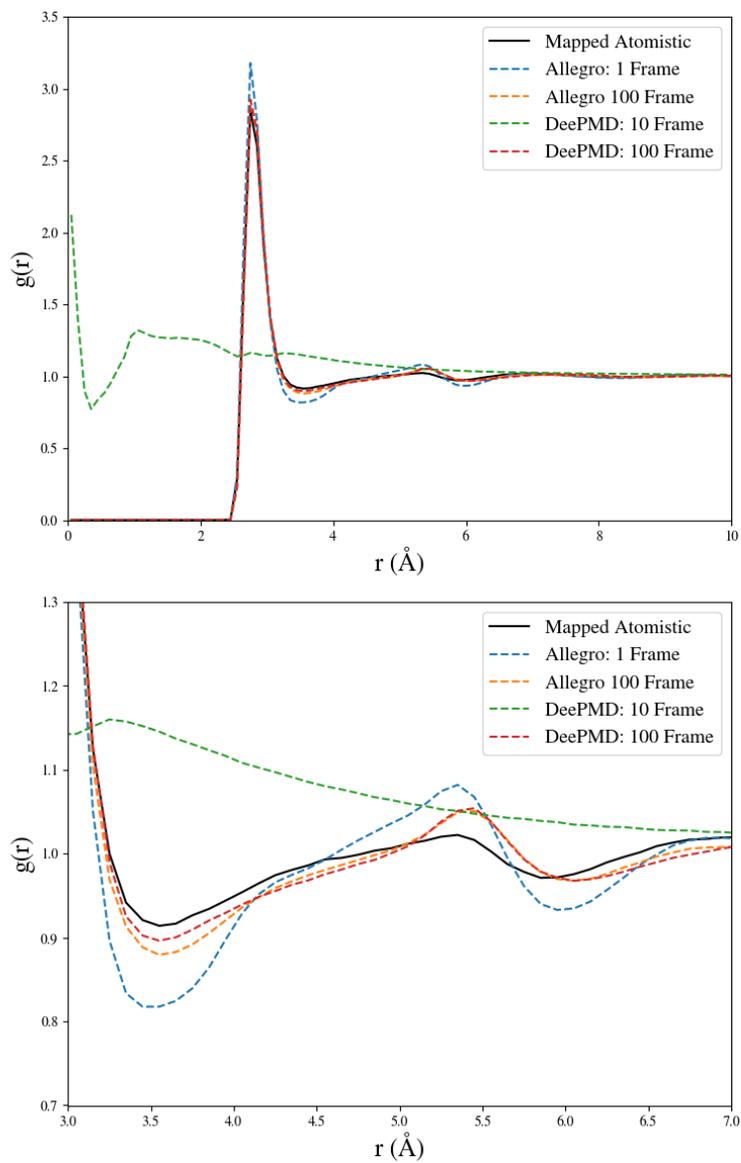

Figure 2. a) Radial distribution functions of CG water for Allegro and DeePMD models compared to reference atomistic data for TIP3P water. b) Detail of RDF in top panel between 3 and 7 Angstroms.

While the 1-frame Allegro model for both TIP3P and SPC/E water is the least accurate of the stable models, it still qualitatively captures the shape of the RDF and, in multiple cases, is slightly more accurate where the peaks and wells are located. Unsurprisingly, there is an overall trend of increasing quality as the number of training examples increases, as seen in both the Allegro and DeePMD models.

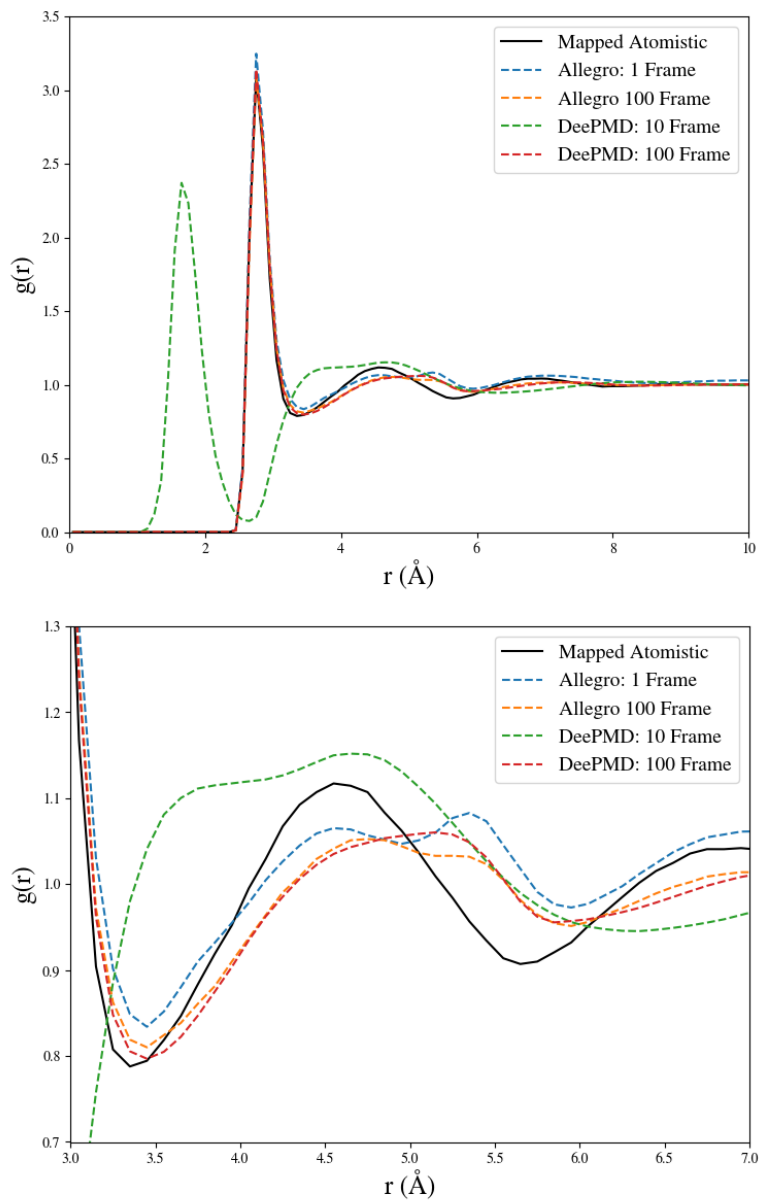

Figure 3. a) Radial distribution functions of CG water for Allegro and DeePMD models compared to reference atomistic data for SPC/E water. b) Detail of RDF in top panel between 3 and 7 Angstroms.

*Three-body correlations*. Each model aside from the 10-frame DeePMD model performs reasonably well by this benchmark, although this is not very surprising given how expressive a DNN-based force field can be. One value of a DNN-based CG force field may be its ability to better capture many-body correlations which can be difficult for pairwise CG potentials. Figures 4 and 5 show water-water-water triplet angular distributions for each model parameterized in comparison to the mapped atomistic TIP3P and SPC/E water models. As with the behavior seen in the RDFs, there is an overall trend of increasing accuracy with respect to increased amounts of training data.

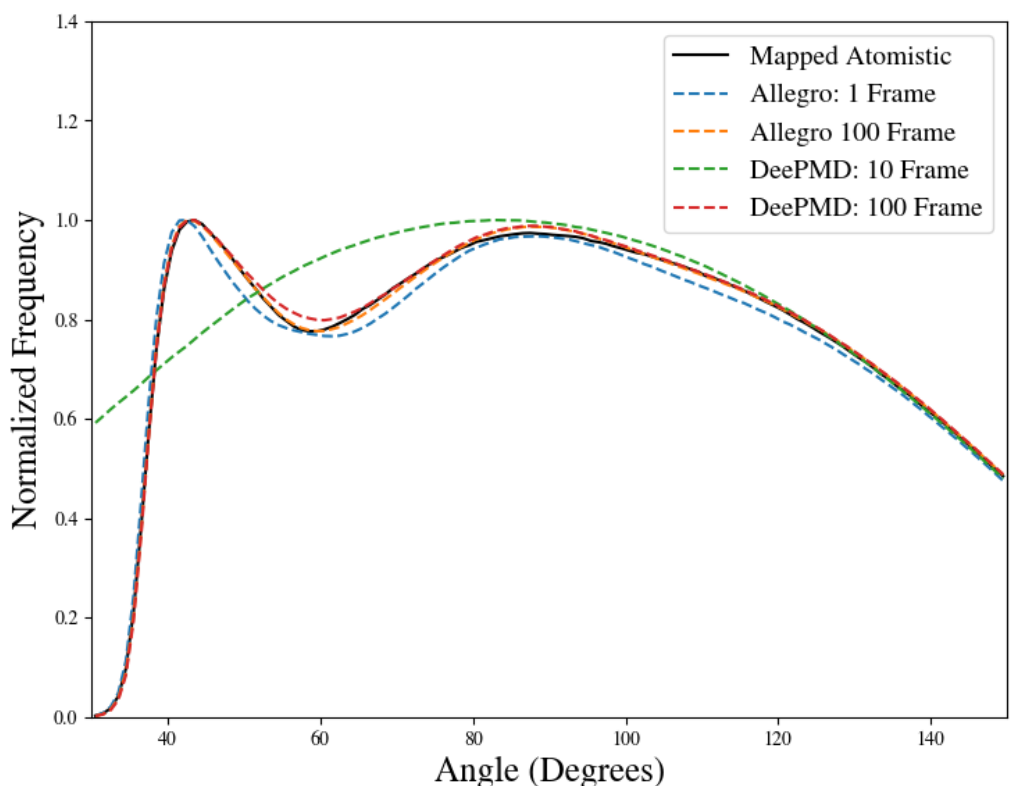

Figure 4. Three body angular distributions between triplets of CG waters for each Allegro and DeePMD model compared to mapped atomistic reference data of TIP3P water. Distributions were calculated using Equation 5.

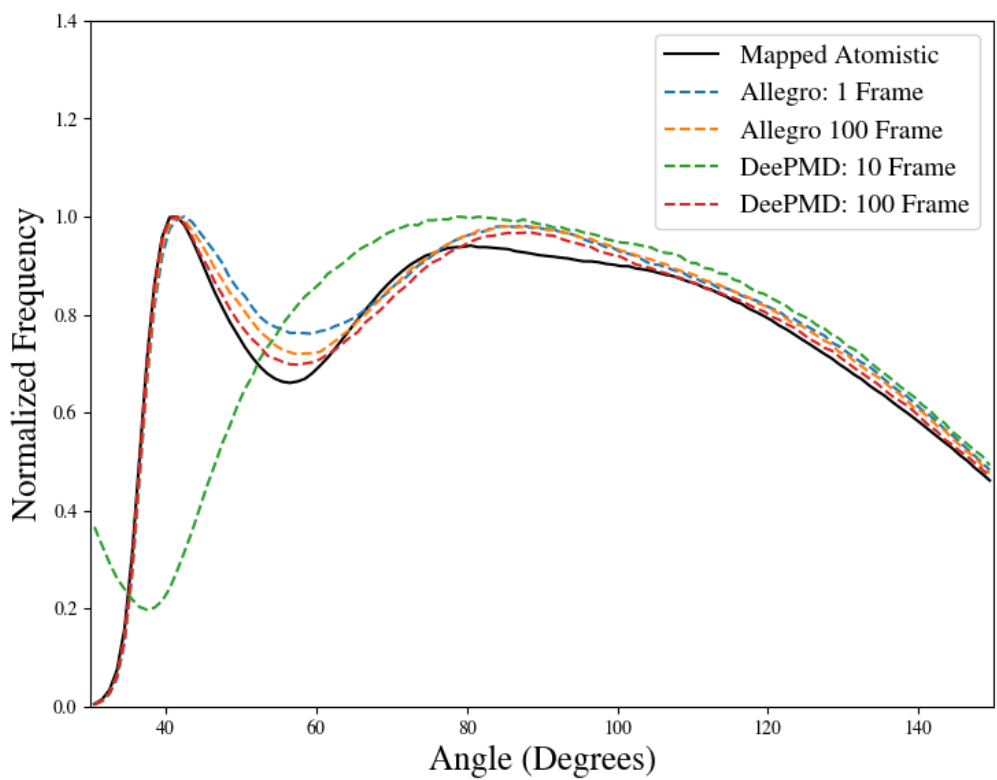

Figure 5. Three body angular distributions between triplets of CG waters for each Allegro and DeePMD model compared to mapped atomistic reference data of SPC/E water. Distributions were calculated using Equation 5.

Between the models trained on 100 frames of data, the Allegro model outperforms DeePMD for the TIP3P AA water model, with a better representation of the minimum around 60 degrees and comparable accuracy everywhere else. For the SPC/E water model, the 100-frame DeePMD water performs slightly better in capturing three-body correlations around the minimum near 60 degrees. Of the stable models, the 1-frame Allegro model is again the least accurate, predicting much lower probability values everywhere except for the first peak at 45 degrees. As before, the 10-frame DeePMD model completely fails to capture the 3-body correlations of both TIP3P and SPC/E water.

While structural correlations have been investigated in this manuscript, we note that dynamical behavior of ML CG models, including diffusion, has not to our knowledge been thoroughly investigated. Such investigations are hindered by the artificial speedup in dynamical behavior exhibited by Newtonian mechanics of the CG PMF compared to the corresponding AA system.[55] This acceleration of CG dynamics is due to the absence of fluctuation and dissipation forces.[56, 57] ML force-fields for AA models should ideally perfectly reproduce the dynamical behavior of the reference model.[58] However, ML CG models will ideally match the dynamical behavior of the PMF, whose value is not knowable without an explicit representation of the PMF in the first place. Consequently, we suggest systematic investigation of the convergence of dynamical behavior of ML CG force-fields is paramount, and a logical route forward from this manuscript.

## 4. Conclusions

In this work, we compare symmetry invariant and symmetry equivariant neural networks in their capacity to generate accurate CG force fields in the limit of low training data. Two architectures, DeePMD and Allegro, were chosen for symmetry invariance and equivariance, respectively. We show that symmetry equivariant models can form stable CG water models with even just a single frame of reference condensed phase MD data. It is shown consistently across both AA water models that the Allegro architecture is more data efficient than that of DeePMD.

It is not surprising that when holding model architecture constant, the models trained on more reference data outperformed those parameterized with less. In all cases, the 100 frame models were able to accurately capture RDFs and 3-body correlations, though there is certainly room for improvement in both the Allegro and DeePMD models. Furthermore, it would be worth investigating how similar architectures such as Schnet perform against these architectures. These architectures could likely produce even better models with the same training datasets if more hyperparameter sweeping was performed. In particular, the fitting and embedding networks were chosen to be far smaller than those used in previous studies to generate atomistic force fields from quantum mechanical simulations. For example, the original Allegro models utilized fitting networks with three hidden layers of 1024 neurons each, instead of the 32-width network used in the current work. This choice was made to maximize their speed as CG models depend on integration speed to further enhance their sampling of the system. Despite this, these models numerically integrate slower than even a corresponding AA system, with the fastest model, Allegro, integrating at a speed of ~25 ns/day on a small 3916 particle system, even when utilizing 4 GPUs. In contrast, simulation of the corresponding AA system in GROMACS integrates at a speed of ~717 ns/day using 4 GPUs. However, it must be noted that the integration time of the CG model cannot be directly compared to the integration time of the AA model because the "time" in the CG model is not same as the physical time of the AA model. For example, a one CG bead water model typically has a diffusion constant 5-10 times larger than the underlying AA system at 300K.

While there is room for improvement in the integration speed of DNN-based CG force fields, the addition of equivariant embedding can reduce the amount of training data required to generate a stable model by orders of magnitude. Though it was not the most accurate model, Allegro could train a force field that reproduced all qualitative features of the 2- and 3-body correlations of water, with a reasonable amount of quantitative accuracy, even when using a single frame of MD training data containing 512 total training examples. In comparison, DeePMD could not create a model which stably formed a bulk liquid using 10x the amount of training data. This sidesteps one of the biggest hurdles for generating DNN CG force fields and suggests that incorporating physical intuition and restraints may increase their training efficiency. With additional advances in DNN integration speed, methods such as these could become the state of the art for CG modeling in the future. Explicit inclusion of bonded CG beads could also expand the capacity of these models to much more complicated systems for which traditional CG methods fail.

**Supporting Information**

Example input files for training of DeePMD and Allegro models

**Acknowledgments**

This material is based upon work supported by the National Science Foundation (NSF Grant CHE-2102677). Simulations were performed using computing resources provided by the University of Chicago Research Computing Center (RCC).

**Data Availability**

The data that support the findings of this work are available from the corresponding author upon request.

# TOC Graphic

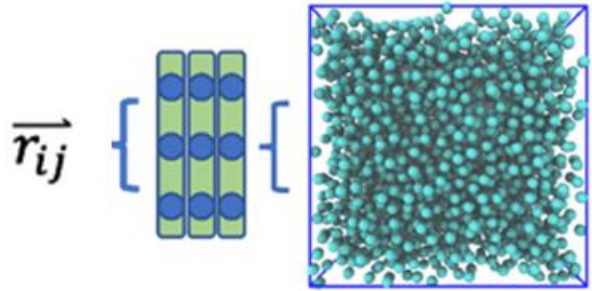

# SUPPORTING INFORMATION
# for

**Coarse-Graining with Equivariant Neural Networks: A Path Towards Accurate and Data-Efficient Models**

Timothy D. Loose,# Patrick G. Sahrmann,# Thomas S. Qu, and Gregory A. Voth*

Department of Chemistry, Chicago Center for Theoretical Chemistry, James Franck Institute, and Institute for Biophysical Dynamics, The University of Chicago, Chicago, IL 60637, USA

#Authors contributed equally

*Corresponding Author: gavoth@uchicago.edu

**Script 1**. Example Input File for DeePMD Model Training.
```
{
    "_comment": " model parameters",
    "model": {
  "_comment": " H and O will be 0 and 1, consistent with type.raw ",
    "type_map":    ["O"],
    "descriptor" :{
        "type":         "se_e2_a",
        "sel":          [60],
        "rcut_smth":    1.00,
        "rcut":         7.00,
        "neuron": [8,16,32],
        "resnet_dt":    false,
        "seed":         1,
        "_comment":     " that's all"
        },
    "fitting_net" : {
        "neuron":       [32,32,32],
        "resnet_dt":    true,
        "seed":         1,
        "_comment":         " that's all"
        },
    "_comment": " that's all"
    },

    "learning_rate" :{
     "type":         "exp",
```

```
      "decay_steps": 5000,
      "start_lr":    0.005,
      "stop_lr":     1.76e-7,
      "_comment":    "that's all"
    },

    "loss" :{
     "start_pref_e":     0,
     "limit_pref_e":     0,
     "start_pref_f":     1000,
     "limit_pref_f":     1000,
     "start_pref_v":     0,
     "limit_pref_v":     0,
     "_comment":     " that's all"
    },

    "_comment": " traing controls",
    "training" : {
     "systems":      ["../data/"],
     "set_prefix":  "set",
     "stop_batch":  100000,
     "batch_size":  [10],

     "seed":         1,

     "_comment": " display and restart",
     "_comment": " frequencies counted in batch",
     "disp_file":   "lcurve.out",
     "disp_freq":   100,
     "numb_test":   10,
     "save_freq":   500,
     "save_ckpt":   "model.ckpt",
     "disp_training":true,
     "time_training":true,
     "profiling":   false,
     "profiling_file":"timeline.json",
     "_comment":    "that's all"
    },

    "_comment":          "that's all"
}
```
**Script 2**. Example Input File for Allegro Model Training.

```
# This file serves as a starting example input file for Allegro
# For a full, detailed set of general training+dataset options see
configs/full.yaml in the NequIP repo:
# https://github.com/mir-group/nequip/blob/main/configs/full.yaml
```

```yaml
# This file additionally documents the Allegro-specific options

# general

# Two folders will be used during the training: 'root'/process and
'root'/'run_name'
# run_name contains logfiles and saved models
# process contains processed data sets
# if 'root'/'run_name' exists, 'root'/'run_name'_'year'-'month'-
'day'-'hour'-'min'-'s' will be used instead.
root: results/water
run_name: CG_water

# model initialization seed
seed: 123456

# data set seed, determines which data to sample from file
dataset_seed: 123456

# set true if a restarted run should append to the previous log
file
append: true

# type of float to use, e.g. float32 and float64
default_dtype: float32

# -- network --
# tell nequip which modules to build
model_builders:
 - allegro.model.Allegro
 # the typical model builders from `nequip` can still be used:
 - PerSpeciesRescale
 - ForceOutput
 - RescaleEnergyEtc

# radial cutoff in length units
r_max: 7.0

# average number of neighbors in an environment is used to
normalize the sum, auto precomputed it automitcally
avg_num_neighbors: auto

# radial basis
# set true to train the bessel roots
BesselBasis_trainable: true
```

```yaml
# p-parameter in envelope function, as proposed in Klicpera, J. et al., arXiv:2003.03123
# sets it BOTH for the RadialBasisProjection AND the Allegro_Module
PolynomialCutoff_p: 6

# symmetry
# maximum order l to use in spherical harmonics embedding, 1 is basedline (fast), 2 is more accurate, but slower, 3 highly accurate but slow
l_max: 1

# whether to include E(3)-symmetry / parity
# allowed: o3_full, o3_restricted, so3
parity: o3_full

# number of tensor product layers, 1-3 usually best, more is more accurate but slower
num_layers: 1

# number of features, more is more accurate but slower, 1, 4, 8, 16, 64, 128 are good options to try depending on data set
env_embed_multiplicity: 2

# whether or not to embed the initial edge, true often works best
embed_initial_edge: true

# hidden layer dimensions of the 2-body embedding MLP
two_body_latent_mlp_latent_dimensions: [8, 16, 32]
# nonlinearity used in the 2-body embedding MLP
two_body_latent_mlp_nonlinearity: silu
# weight initialization of the 2-body embedding MLP
two_body_latent_mlp_initialization: uniform

# hidden layer dimensions of the latent MLP
latent_mlp_latent_dimensions: [32,32,32]

# nonlinearity used in the latent MLP
latent_mlp_nonlinearity: silu

# weight initialization of the latent MLP
latent_mlp_initialization: uniform

# whether to use a resnet update in the scalar latent latent space, true works best usually
latent_resnet: true
```

```yaml
# hidden layer dimensions of the environment embedding mlp, none work best (will build a single linear layer)
env_embed_mlp_latent_dimensions: []

# nonlinearity used in the environment embedding mlp
env_embed_mlp_nonlinearity: null

# weight initialzation of the environment embedding mlp
env_embed_mlp_initialization: uniform

# - end allegro layers -

# Final MLP to go from Allegro latent space to edge energies:

# hidden layer dimensions of the per-edge energy final MLP
edge_eng_mlp_latent_dimensions: [8]

# nonlinearity used in the per-edge energy final MLP
edge_eng_mlp_nonlinearity: null

# weight initialzation in the per-edge energy final MLP
edge_eng_mlp_initialization: uniform

# -- data --
# there are two options to specify a dataset, npz or ase
# npz works with npz files, ase can ready any format that ase.io.read can read
# in most cases working with the ase option and an extxyz file is by far the simplest way to do it and we strongly recommend using this
# simply provide a single extxyz file that contains the structures together with energies and forces (generated with ase.io.write(atoms, format='extxyz', append=True))
# for a simple snippet to do this, see the gists here: https://github.com/simonbatzner

# npz option
dataset:                                                    ase
# type of data set, can be npz or ase
dataset_file_name: ./water.lammpstrj                        # path to data set file
key_mapping:
  z:                                              atomic_numbers
# atomic species, integers
  E:                                                total_energy
# total potential eneriges to train to
```

```yaml
    F:                                                      forces
# atomic forces to train to
    R:                                                      pos
# raw atomic positions
npz_fixed_field_keys:
# fields that are repeated across different examples
  - atomic_numbers

# ase option
# dataset: ase
# dataset_file_name: filename.extxyz
# ase_args:
#    format: extxyz

# A mapping of chemical species to type indexes is necessary if
the dataset is provided with atomic numbers instead of type
indexes.
chemical_symbols:
  - H

# logging
# whether to use weight and biases (see wandb.ai)
wandb: false

# project name in wandb

# the same as python logging, e.g. warning, info, debug, error.
case insensitive
verbose: debug

# training

# number of training samples to use
n_train: 10

# number of validation samples to use
n_val: 50

# batch size, we found it important to keep this small for most
applications including forces (1-5); for energy-only training,
higher batch sizes work better
batch_size: 1

# stop training after _ number of epochs, we set a very large
number here, it won't take this long in practice and we will use
early stopping instead
max_epochs: 1000000
```

```yaml
# learning rate, we found values between 0.002 and 0.0005 to work
best - this is often one of the most important hyperparameters to
tune
learning_rate: 0.002

# can be random or sequential. if sequential, first n_train
elements are training, next n_val are val, else random, usually
random is the right choice
train_val_split: random

# If true, the data loader will shuffle the data, almost always a
good idea
shuffle: true

# metrics used for scheduling and saving best model. Options:
`set`_`quantity`, set can be either "train" or "validation,
"quantity" can be loss or anything that appears in the validation
batch step header, such as f_mae, f_rmse, e_mae, e_rmse
metrics_key: validation_loss

# use an exponential moving average of the weights
# if true, use exponential moving average on weights for val/test,
usually helps a lot with training, in particular for energy errors
use_ema: true

# ema weight, typically set to 0.99 or 0.999
ema_decay: 0.99

# whether to use number of updates when computing averages
ema_use_num_updates: true

# loss function
# different weights to use in a weighted loss functions
# if you use peratommseloss, then this is already in a per-atom
normalized space (both E/F are per-atom quantities)
# in that case, 1:1 works best usually
loss_coeffs:
  forces: 1.
  total_energy:
    - 1.
    - PerAtomMSELoss

# optimizer
# default optimizer is Adam
optimizer_name: Adam
optimizer_params:
```

```yaml
  amsgrad: false
  betas: !!python/tuple
  - 0.9
  - 0.999
  eps: 1.0e-08
  weight_decay: 0.

# lr scheduler, drop lr if no improvement for 50 epochs
# on-plateau, reduce lr by factory of lr_scheduler_factor if metrics_key hasn't improved for lr_scheduler_patience epoch
lr_scheduler_name: ReduceLROnPlateau
lr_scheduler_patience: 50
lr_scheduler_factor: 0.5

# early stopping if max 7 days is reached or lr drops below 1e-5 or no improvement on val loss for 100 epochs
early_stopping_upper_bounds:
  cumulative_wall: 604800.

early_stopping_lower_bounds:
  LR: 1.0e-5

early_stopping_patiences:
  validation_loss: 100
```